\documentclass{PoS}

\usepackage{amssymb}
\usepackage{amsthm}
\usepackage{amsmath}
\usepackage{graphicx}
\usepackage{subcaption}

\newcommand{\be}{\begin{eqnarray}}
\newcommand{\ee}{\end{eqnarray}}
\newcommand{\ma}{\mathrm}
\newcommand{\ml}{\mathcal}
\newcommand{\bs}{\boldsymbol}

\title{Coupled Transport Equations for Quarkonium Production in Heavy Ion Collisions}
\ShortTitle{}

\author{\speaker{Xiaojun Yao}\\
        Center for Theoretical Physics, Massachusetts Institute of Technology, \\
        Cambridge, MA, 02139, USA}

\author{{Weiyao Ke}\\
	Nuclear Science Division, Lawrence Berkeley National Laboratory, \\
Berkeley, CA 94720, USA}

\author{{Yingru Xu}, {Steffen A. Bass} and {Berndt M\"uller}\\
	Department of Physics, Duke University, \\
Durham, NC 27708, USA}

\abstract{Motivated by recent applications of the open quantum system formalism to understand quarkonium transport in the quark-gluon plasma, we develop a set of coupled Boltzmann equations for open heavy quark-antiquark pairs and quarkonia. Our approach keeps track of the correlation between the heavy quark-antiquark pair from quarkonium dissociation and thus is able to account for both uncorrelated and correlated recombination. By solving the coupled Boltzmann equations for current heavy ion collision experiments, we find correlated recombination is crucial to describe the data of bottomonia nuclear modification factors. To further test the importance of correlated recombination in experiments, we propose a new observable: $\frac{R_{AA}[\chi_b(1P)]}{R_{AA}[\Upsilon(2S)]}$. Future measurements of this ratio will help distinguish calculations with and without correlated recombination.}

\FullConference{%
  HardProbes2020\\
  1-6 June 2020\\
  Austin, Texas}

\begin{document}

\section{Introduction}
Quarkonium suppression in heavy ion collisions has been studied for many years as a tool to probe the quark-gluon plasma (QGP). Both cold and hot nuclear matter effects contribute to the suppression. The shadowing of the parton distribution function (PDF) at small Bjorken $x$ in the heavy nucleus is one example of the cold nuclear matter effects. Hot medium effects include screening of the heavy quark-antiquark ($Q\bar{Q}$) potential, dissociation of quarkonium in dynamical scattering processes, and recombination of unbound $Q\bar{Q}$ pairs. Many studies applied semiclassical transport equations that account for these three hot medium effects to describe quarkonium evolution in the QGP, and achieved phenomenological success.

Recent developments based on the open quantum system formalism provide new insights about quarkonium evolution in the QGP~\cite{Akamatsu:2011se,Blaizot:2018oev,Yao:2018nmy,Brambilla:2019tpt,Miura:2019ssi,Yao:2020eqy}. The $Q\bar{Q}$ pairs interacting with the QGP can be treated as an open quantum system and its evolution is governed by the Lindblad equation in the weak coupling limit. The Lindblad equation can be recast as a stochastic Schr\"odinger equation, in which random forces distort the wavefunction during the evolution. As a result, the wavefunction of the $Q\bar{Q}$ pair loses coherence as time goes by~\cite{Akamatsu:2011se}. For example, if the wavefunction is the $1S$ bound state at $t=0$, $|\psi(t=0)\rangle = |1S\rangle $ (we assume a constant temperature here for illustration of the key idea), the wavefunction decoherence will lead to $|\langle 1S| \psi(t)\rangle|^2 < 1$, which is interpreted as the dissociation of the $1S$ bound state. But at the same time, if the $2S$ state exists in the medium (i.e., the local medium temperature is below the melting temperature of the $2S$ state), we will also have $|\langle 2S| \psi(t)\rangle|^2 > 0$, i.e., the $2S$ state is regenerated from the dissociating $1S$ state. This type of recombination is different from the traditional recombination studied by the community since the recombining $Q\bar{Q}$ pair in the former case originates from a dissociating quarkonium, while the pair in the latter mostly comes from uncorrelated heavy quarks produced from different initial hard vertices. To distinguish these two cases, the former recombination is named correlated recombination while the latter uncorrelated recombination. In a nutshell, quarkonium dissociation is a result of the wavefunction decoherence, which also leads to correlated recombination at the same time.

Phenomenological consequence of uncorrelated recombination has been investigated in both transport equations~\cite{Thews:2000rj,Grandchamp:2003uw,Yao:2017fuc} and statistical hadronization models~\cite{Andronic:2007bi}. But few studies explored the phenomenological consequence of correlated recombination. To this end, one can solve the Lindblad equation or the stochastic Schr\"odinger equation with a realistic hydrodynamic description of the QGP in heavy ion collisions. On the other hand, one can try to incorporate correlated recombination into semiclassical transport calculations, which requires keeping track of the correlation between the $Q\bar{Q}$ pair from the quarkonium dissociation. This motivates us to construct the coupled Boltzmann equations for open $Q\bar{Q}$ pairs and quarkonia. We will briefly explain the coupled Boltzmann equations in Section~\ref{sect:transport} and discuss the phenomenological implication of correlated recombination in Section~\ref{sect:results}. A new experimental observable will be proposed in Section~\ref{sect:new}, which we believe can further test the importance of correlated recombination. Finally in Section~\ref{sect:conclusion}, conclusions are drawn.

\section{Coupled Boltzmann Equations for Heavy Flavors}
\label{sect:transport}
The set of coupled Boltzmann transport equations for unbound heavy $Q\bar{Q}$ pairs and quarkonia is given by
\be
\label{eq:LBE}
(\frac{\partial}{\partial t} + \dot{{\bs x}} _Q\cdot \nabla_{{\bs x}_Q} + \dot{{\bs x}} _{\bar{Q}}\cdot \nabla_{{\bs x}_{\bar{Q}}} ) f_{Q\bar{Q}}({\bs x}_Q, {\bs p}_Q, {\bs x}_{\bar{Q}}, {\bs p}_{\bar{Q}}, t) &=& \ml{C}_{Q\bar{Q}}  -  \ml{C}_{Q\bar{Q}}^{+} +  \ml{C}_{Q\bar{Q}}^{-}\\
(\frac{\partial}{\partial t} + \dot{{\bs x}}\cdot \nabla_{\bs x})f_{nls}({\bs x}, {\bs p}, t) &=& \ml{C}_{nls}^{+}-\ml{C}_{nls}^{-}\,,
\ee
where $f$ denotes the phase space distribution. Each quarkonium state is labeled by their quantum number $nls$ ($n$ is for the radial excitation, $l$ the orbital angular momentum and $s$ the spin). On the right hand sides, collision terms with a superscript ``+(-)" represent the contribution from recombination (dissociation). The collision term without superscripts $\ml{C}_{Q\bar{Q}}$ denotes the processes where the $Q\bar{Q}$ pairs exchange energy and momentum with the QGP. If we neglect the interaction between the heavy quark-antiquark pair, we can write
\be
\label{eqn:collision_HQ}
\ml{C}_{Q\bar{Q}} = \ml{C}_{Q} + \ml{C}_{\bar{Q}}\,,
\ee
i.e., the heavy quark and antiquark interact independently with the medium. This is probably a valid assumption because of the screening of the $Q\bar{Q}$ potential and the average over color. A factorized $Q\bar{Q}$ distribution $f_{Q\bar{Q}}({\bs x}_Q, {\bs p}_Q, {\bs x}_{\bar{Q}}, {\bs p}_{\bar{Q}}, t) = f_{Q}({\bs x}_Q, {\bs p}_Q, t) f_{\bar{Q}}({\bs x}_{\bar{Q}}, {\bs p}_{\bar{Q}}, t)$ will lead to Eq.~(\ref{eqn:collision_HQ}). But in general Eq.~(\ref{eqn:collision_HQ}) does not imply a factorized distribution. Many model calculations of recombination implicitly assume the factorization of the $Q\bar{Q}$ distribution and thus fail to account for the correlated recombination. Detailed expressions of the collision terms can be found in Refs.~\cite{Yao:2020xzw,Yao:2018sgn}.

\section{Results}
\label{sect:results}
We solve the coupled transport equations by Monte Carlo simulations. The initial distributions are sampled from \textsc{Pythia} \cite{Sjostrand:2014zea} for momenta and {T\raisebox{-0.5ex}{R}ENTo} \cite{Moreland:2014oya} for positions. We use the nuclear parton distribution function (nPDF) parametrized by EPPS16 \cite{Eskola:2016oht}. A calibrated $2+1$D viscous hydrodynamics is applied to describe the bulk dynamics \cite{Song:2007ux,Shen:2014vra,Bernhard:2016tnd}. Details of the simulations can be found in Ref.~\cite{Yao:2020xzw}.

\begin{figure}
    \centering
    \begin{subfigure}[t]{0.45\textwidth}
        \centering
        \includegraphics[height=2.0in]{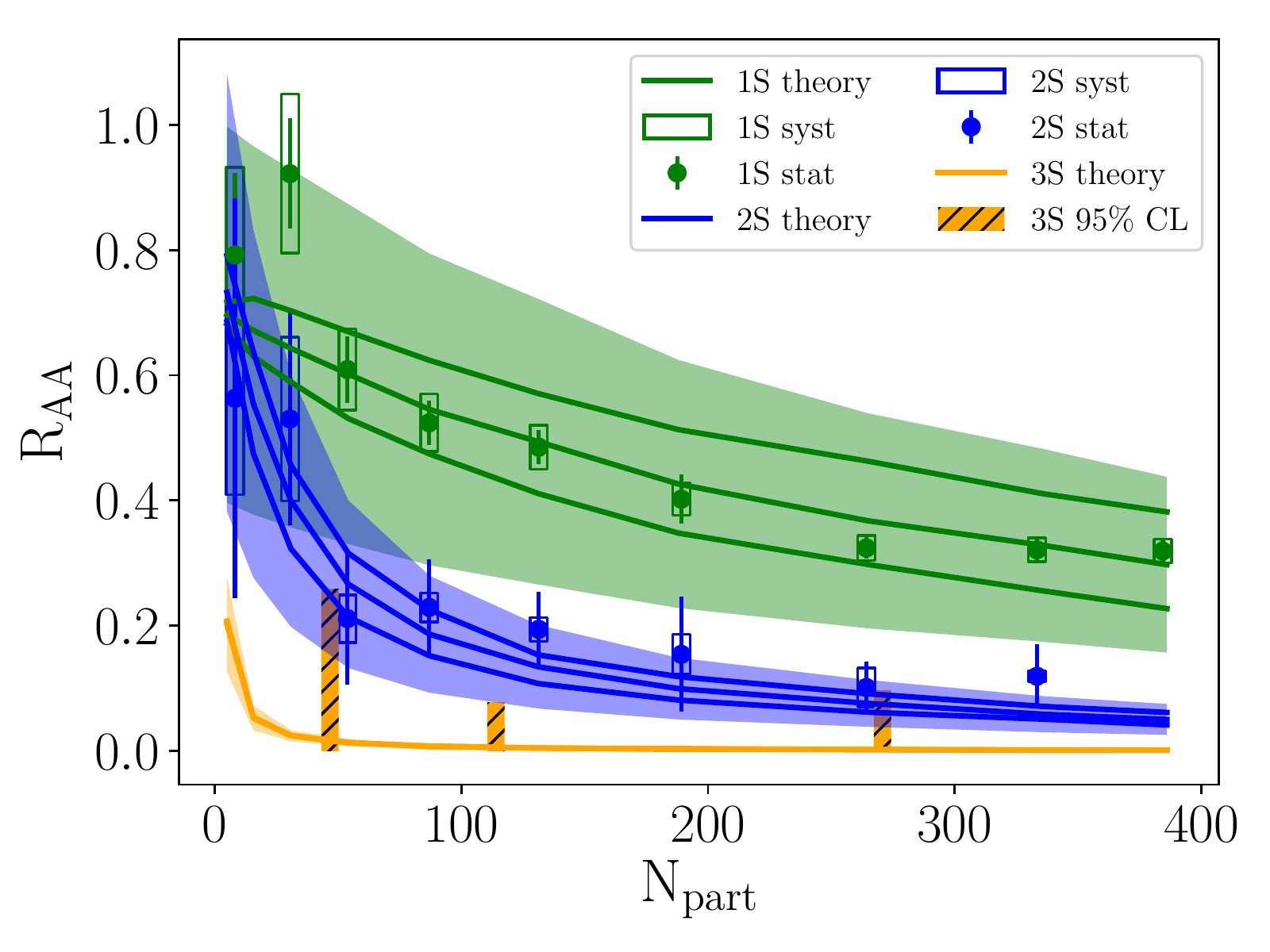}
        \caption{With cross-talk recombination.}
        \label{fig:502Npart_corr}
    \end{subfigure}%
    ~
    \begin{subfigure}[t]{0.45\textwidth}
        \centering
        \includegraphics[height=2.0in]{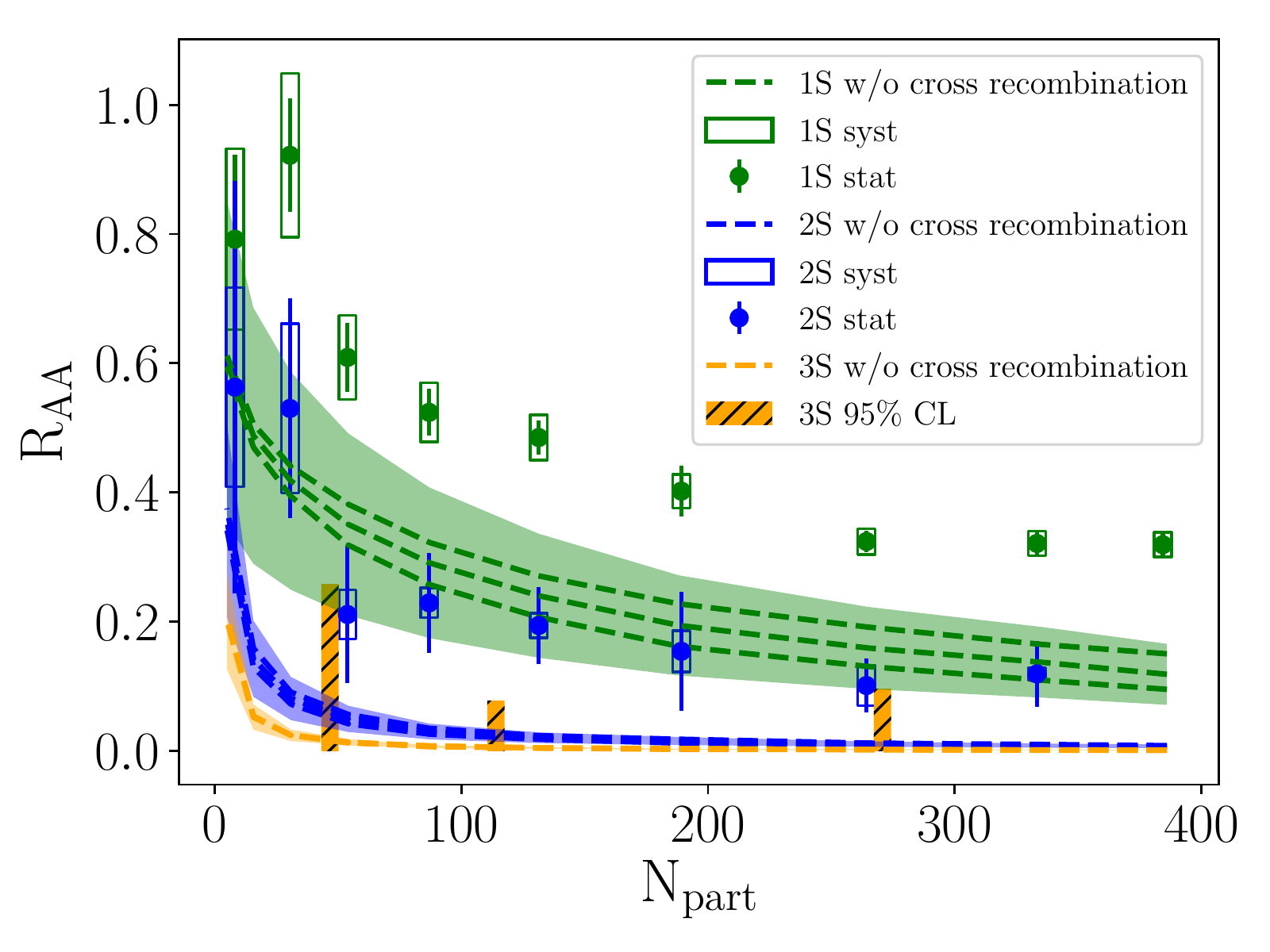}
        \caption{Without cross-talk recombination.}
        \label{fig:502Npart_nocorr}
    \end{subfigure}
    \caption{Bottomonia $R_\ma{AA}$ as functions of centrality at $5.02$ TeV Pb-Pb collisions. Experimental data are taken from Ref.~\cite{Sirunyan:2018nsz}.}
    \label{fig:502Npart}
\end{figure}

Instead of showing the full comparison between our calculation results and experimental data, we focus on discussing the importance of correlated recombination here. In Fig.~\ref{fig:502Npart}, we show the results of the nuclear modification factor $R_{AA}$ of $\Upsilon(nS)$ as a function of centrality, compared with the CMS measurements. The three solid curves in the plots correspond to three different choices of two model parameters. The parameters used for the upper and lower curves differ by $10\%$ from the middle curve. The filled band represents the uncertainty from nPDF. The nPDF uncertainties in the EPPS16 parametrization dominate over the model parameter uncertainties. In the right plot, correlated cross-talk recombination is turned off (cross-talk means $2S\to1S$ for example). It can be seen that correlated recombination is crucial to describe the data,  even if we take into account the uncertainties from the parameters and the nPDF.

\begin{figure}
    \centering
    \begin{subfigure}[t]{0.45\textwidth}
        \centering
        \includegraphics[height=2.0in]{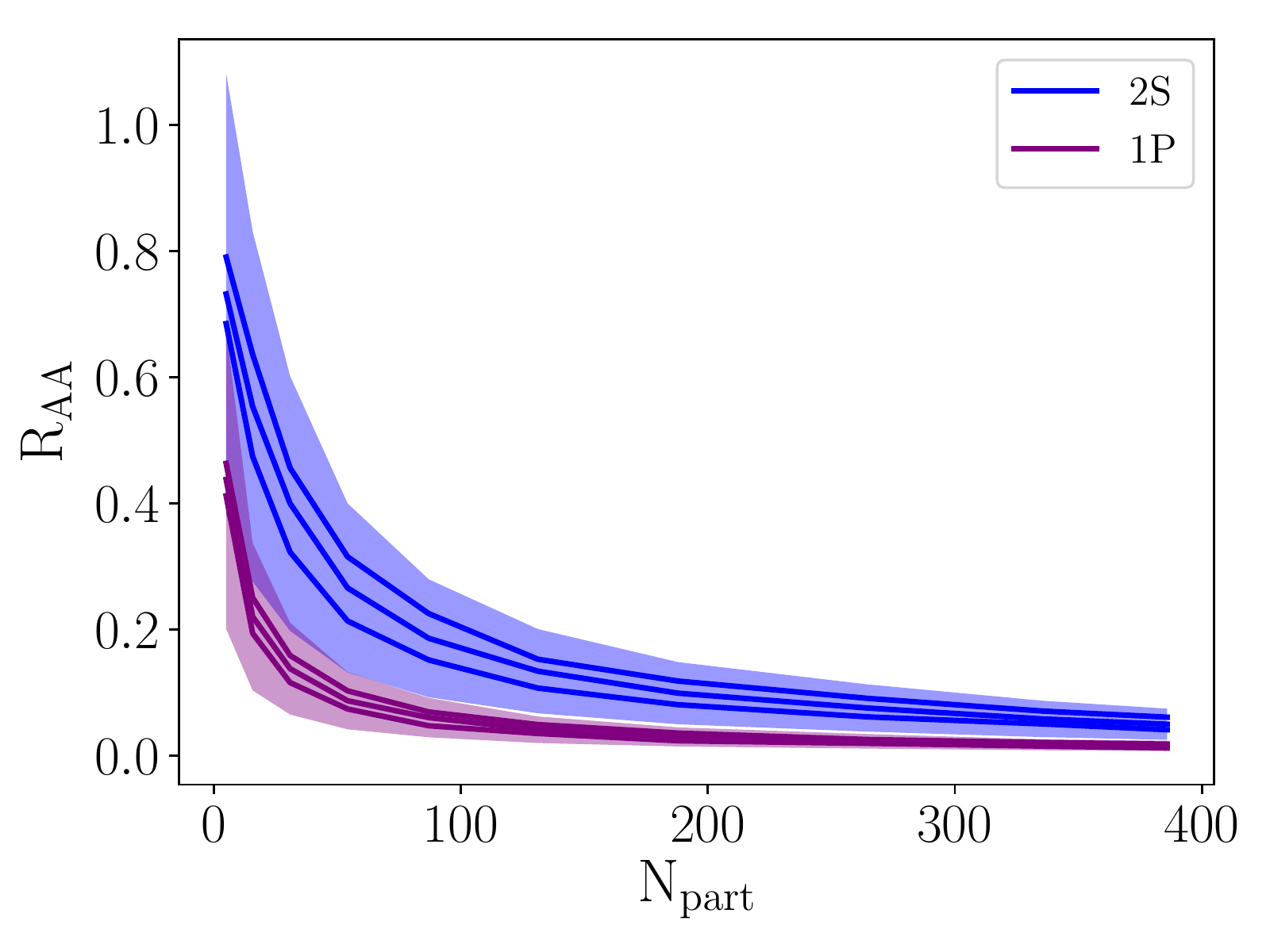}
        \caption{With cross-talk recombination.}
    \end{subfigure}%
    ~
    \centering
    \begin{subfigure}[t]{0.45\textwidth}
        \centering
        \includegraphics[height=2.0in]{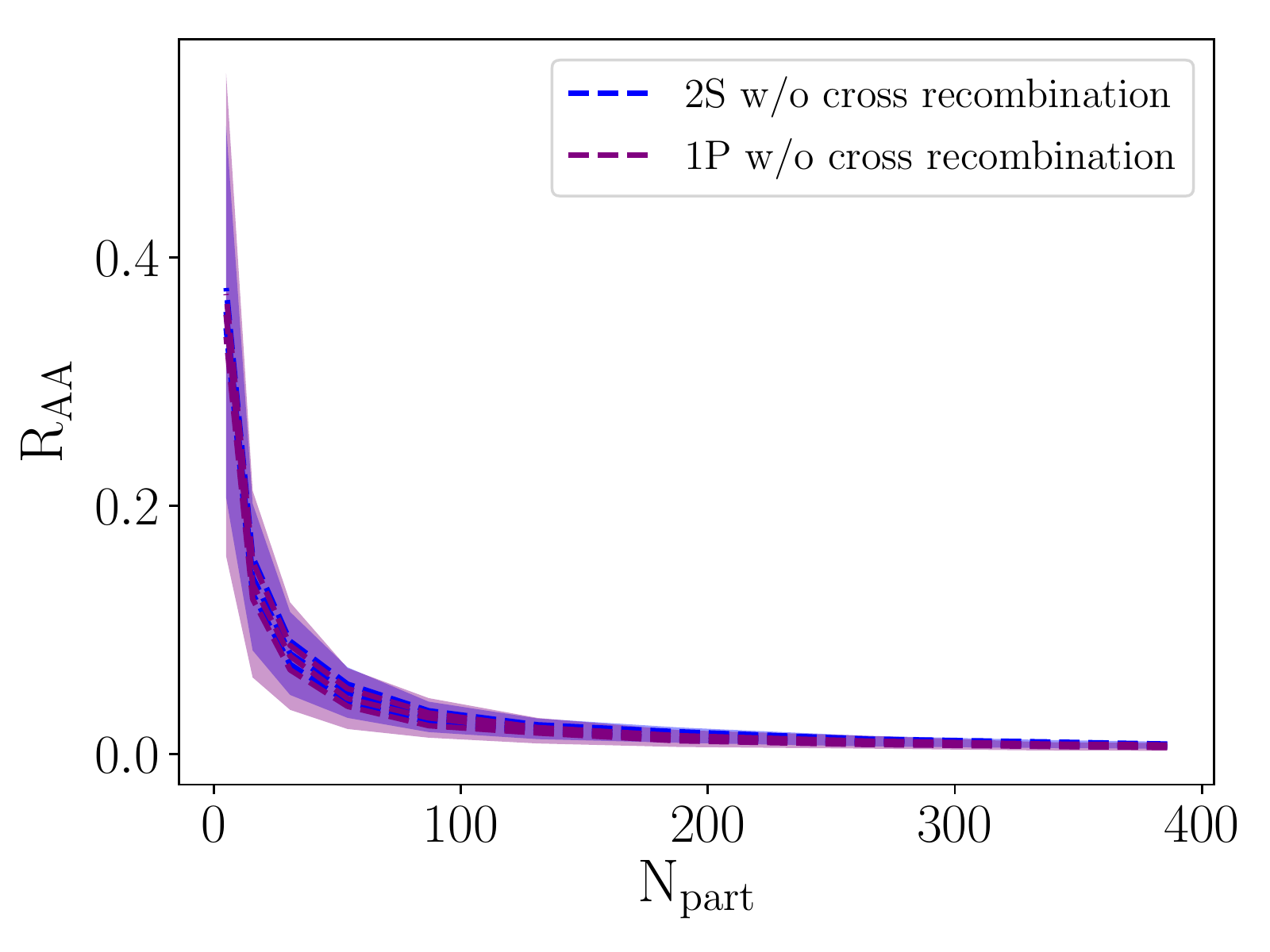}
        \caption{Without cross-talk recombination.}
    \end{subfigure}%
\caption{$R_{\ma{AA}}(\chi_b(1P))$ compared with $R_{\ma{AA}}(\Upsilon(2S))$.}
\label{fig:chi_b}
\end{figure}

\section{New Observable}
\label{sect:new}
As discussed in the previous section, correlated recombination is crucial to describe the data. One may argue that similar agreement with the data can be achieved by tuning the parameters in calculations without correlated recombination added. It is not easy to distinguish the two scenarios by using currently available experimental data. To demonstrate the importance of correlated recombination, we propose a new observable: $\frac{R_{AA}[\chi_b(1P)]}{R_{AA}[\Upsilon(2S)]}$, which will be dramatically different between calculations with and without correlated recombination. Results of this new observable in our calculations are shown in Fig.~\ref{fig:chi_b}. Without correlated recombination, $R_{AA}[\chi_b(1P)]$ and $R_{AA}[\Upsilon(2S)]$ are close to each other since the binding energies of the $1P$ and $2S$ states are close. So the ratio is about one, which can also be seen in results of Ref.~\cite{Krouppa:2015yoa} in which the calculation has no correlated recombination. However, once correlated recombination is taken into account, an initial $2S$ state can turn into a $1P$ state via first dissociation and then correlated recombination, and similarly for $1P\to2S$. Since their binding energies are close, these processes have similar rates. But primordially, more $1P$ states are produced than $2S$ states. As a result, more $2S$ states are regenerated from dissociating $1P$ states than $1P$ states regenerated from dissociating $2S$ states. Therefore, the $2S$ production is less suppressed than the $1P$, as shown in the left plot of Fig.~\ref{fig:chi_b}. The ratio is about one third in our calculation. This new observable is powerful in distinguishing models with and without correlated recombination.

\section{Conclusions}
\label{sect:conclusion}
We develop a set of coupled Boltzmann equations for open $Q\bar{Q}$ pairs and quarkonia to explore the phenomenological impact of correlated recombination. Our calculations show correlated recombination is crucial to describe the data. We propose a new experimental observable $\frac{R_{AA}[\chi_b(1P)]}{R_{AA}[\Upsilon(2S)]}$ which is powerful in distinguishing models with and without correlated recombination.

This material is based upon work supported by the U.S. Department of Energy, Office of Science, Office of Nuclear Physics under grant Contract Numbers DE-SC0011090, DE-AC02-05CH11231 and DE-FG02-05ER41367. KW acknowledges support from National Science Foundation under the grant ACI-1550228 within the JETSCAPE Collaboration. XY acknowledges support from Department of Physics, Massachusetts Institute of Technology.

\end{document}